\documentclass[aip,reprint]{revtex4-1}

\usepackage{graphicx}
\usepackage{amsmath}
\usepackage{amssymb}

\usepackage{color}

\makeatletter
\renewcommand{\p@subsection}{}
\renewcommand{\p@subsubsection}{}
\makeatother

\draft 
\begin{document}

\title{Bistable dynamics underlying excitability of ion homeostasis in neuron models} 

\author{Niklas H\"ubel}
\author{Eckehard Sch\"oll}
\affiliation{Department of Theoretical Physics, Technische Universit{\"a}t Berlin, Germany}

\author{Markus A. Dahlem}
\thanks{Corresponding author}
\email[]{dahlem@physik.hu-berlin.de}
\affiliation{Department of Physics, Humboldt Universit\"at zu Berlin, Berlin, Germany}

\date{\today}

\begin{abstract} 
   When neurons fire action potentials, dissipation of free energy is usually not directly considered, because the change in free energy is often negligible compared to the immense reservoir stored in neural transmembrane ion gradients and the long--term energy requirements are met through chemical energy, i.e., metabolism.  However, these gradients can temporarily nearly vanish  in neurological diseases, such as migraine and stroke, and in traumatic brain injury from concussions to severe injuries. We study biophysical neuron models based on the Hodgkin--Huxley (HH) formalism extended to include time--dependent ion concentrations inside and outside the cell and metabolic energy--driven pumps. We reveal the basic mechanism of a state of free energy--starvation (FES) with bifurcation analyses showing that ion dynamics is for a large range of  pump rates bistable  without contact to an ion bath.  This is interpreted as a threshold reduction of a new fundamental mechanism of {\it ionic excitability} that causes a long--lasting but transient FES as observed in  pathological states. We can in particular conclude that a coupling of extracellular ion concentrations to a large glial--vascular bath can take a role as an inhibitory mechanism crucial in ion homeostasis, while the Na$^+$/K$^+$ pumps alone are insufficient to recover from FES. Our results provide the missing link between the HH formalism and activator--inhibitor models that have been successfully used for modeling migraine phenotypes, and therefore will allow us to validate the hypothesis that migraine symptoms are explained by disturbed function in ion channel subunits, Na$^+$/K$^+$ pumps, and other proteins that regulate ion homeostasis.    \end{abstract}

\maketitle

\section*{Author summary}

Theoretical neuroscience complements experimental and clinical neuroscience.
Simulations and analytical insights help to interpret
data and guide our principal understanding of the nervous systems in both
health and disease.  The Hodgkin--Huxley--formulation of action potentials is
certainly one of the most successful models in mathematical biology. It
describes an essential part of cell--to--cell communication in the brain.  This
model was in various ways extended to also describe when the brain's normal
performance fails, such as in migraine hallucinations and acute stroke.
However, the fundamental mechanism of these extensions remained poorly
understood.  We study the structure of biophysical neuron models that starve
from their `free' energy, that is, the energy that can directly be converted to
do work. Although neurons still have access to chemical energy, which needs to
be converted by the metabolism to obtain free energy, their free
energy--starvation can be more stable than expected, explaining pathological
conditions in migraine and stroke.

\section*{Introduction}\label{sec:1}

The Hodgkin--Huxley (HH) model is one of the most successful models in
mathematical biology\cite{HOD52}. This formalism, i.e., a HH--type model,
describes voltage changes across cell membranes that result in excitability.
Not only neurons are excitable cells,  also myocytes, pancreatic
$\beta$--cells, and even a plant cell (Chara corallina) exhibit excitable
dynamics \cite{BEI79,EBI80,CHA83}. The dynamic range of phenomena includes
single action potentials (spikes), periodic spiking, and bursting (slow
modulation of spiking). For example, in pancreatic $\beta$--cells bursting is
induced by a calcium current \cite{ATW80,CHA83}. A more complete treatment of
this phenomenon, however, also requires the inclusion of Na$^+$/K$^+$ pumps
\cite{CHA11}. The dynamics of ion pumps and ion concentrations is
also crucial for cardiac alternans (periodic beat--to--beat variations) and
higher--order rhythms in the ischemic ventricular muscle \cite{DIF85,DOK93,NOB01}.

In the literature such augmented HH--type models are also called
second--generation HH models\cite{ARC00}. In the context of certain pathologies
of the brain, whose fundamental dynamic structure we study here, we prefer the
simpler name `ion--based' models. This indicates that ion concentrations are
major dynamical, that is, time--dependent  variables. Their dynamical role in
neuron models goes beyond merely modulating spiking activity.  Ion dynamics can
lead to a completely new type of {\em ionic excitability}  and bistability,
that is, the phenomena of so--called `spreading depolarizations' and `anoxic
depolarization', respectively. (Spreading depolarizations are also called 'spreading depression'
and we will use both names interchangeably in this paper.) These depolarized states of neurons are related
to migraine, stroke, brain injury, and brain death, that is, to pathologies of
the brain in which a transient or permanent break--down of the transmembrane
potential occurs \cite{DRE11,CHA13a}. Another even more characteristic property
of this `twilight state close to death' \cite{DRE12} are the nearly completely
flat transmembrane ion gradients. The almost complete break--down of both
membrane potential and---due to reduced ion gradients---Nernst potentials together cause a
nearly complete release of the Gibbs free energy, that is, the thermodynamic
potential that measures the energy available to the neurons for normal functioning.  We
hence refer to this state as a state of free energy--starvation (FES).
We want to stress that such phenomena require the broader thermodynamical perspective,
because it goes beyond the HH description 
in terms equivalent electrical circuits in membrane physiology (see discussion).

The object of this study is to clarify quantitatively the detailed ion--based
mechanisms, in particular the time--dependent potentials, leading to this condition. 
In fact, early ion--based models have been introduced in a different context to
describe excitable myocytes and pancreatic $\beta$--cells with variable ion
concentrations \cite{VAR97a,END00,CHA09a}. Neuronal ion--based models have been
used to study spreading depolarizations (SD)
\cite{KAG00,SHA01,MIU07,SOM08,ZAN11,CHA13} and anoxic depolarizations
\cite{ZAN11}. In these phenomenological studies the types of ion dynamics related to the pathologies 
have been reproduced, but not investigated in a bifurcation analysis. Hence the fundamental phase space
structure of these high--dimensional models that underlies the ionic excitability characterisitic of SD
remains poorly understood.  Furthermore,  neuronal ion--based models  have been
used to study  seizure activity \cite{CRE08,BAR11} and spontaneous spiking
patterns in myelinated axons with injury--like membrane damaging conditions
(e.g., caused by concussions) \cite{YU12a,BOU12a}. In these models, the phase
space structure was investigated, however, only with respect to the modulating
effect of ion concentrations on the fast spiking dynamics (seizure activity,
injuries), and with respect to spiking node--to--node transmission fidelity
(myelinated axons).

In this paper we present bifurcation analyses of several minimal biophysical ion--based 
models that reveal bistability of extremely different ion configurations---physiological
conditions vs.\ free energy--starvation---for a large range of pump rates. 
In related models certain bistabilities have been explored before. For example,
Fr\"ohlich et al.\cite{FRO08,FRO06,BAZ04} found coexistence of quiescence and bursting
for certain fixed extracellular potassium concentrations and also bistability
of a physiological and a strongly depolarized membrane state in a slow--fast
a\-na\-ly\-sis of calcium gated channels. Bistability of similar fixed points has
also been found for the variation of extracellular potassium\cite{AIH83} or,
similarly, the potassium Nernst potential\cite{HAH01}. Also the effect of pump strength variation
has been explored under fixed FES conditions\cite{FLO08}. In this paper, however,
we do not treat slow variables as parameters and show bistability of fast
dynamics, but instead we address the stability of ion concentrations
themselves, which are subject to extremely slow dynamics. This allows us to
find bistability of extremely different ion distributions, a feature that
distinguishes these two states from the polarized and depolarized states
studied in the afore mentioned work. A study that also had significantly
different ion distributions was done by Cressman et
al\cite{CRE09}, however, the seizure-like phenomena discussed in their work are
quite different---though clinically related---from those presented in this paper.

Because of the occurrence of ion state bistability we conjecture that our model
describes a threshold reduction of a mechanism that leads to ionic excitability
in form of spreading depolarizations. In other words, we conclude that an
important inhibitory mechanism to describe ion homeostasis such as glial
buffering or diffusive regulation of extracellular ion concentrations plays a
crucial role in ion homeostasis and the Na$^+$/K$^+$ pumps alone are
insufficient to recover from free energy--starved states. We show that when the
extracellular K$^+$ concentration is regulated by linearly coupling it to an
infinite bath, the bistable system changes to an excitable system, which we
call ionic excitability. The effect of turning off glial buffering and
diffusion has been discussed in more detailed ion--based
models\cite{BAZ04,FRO08} before, but has not been related to the fundamental
phase space structure of the system. Our conclusions have been validated by
demonstrating the robustness of the results in a large variety of minimal
ion--based models, which all consistently show this insufficiency of
Na$^+$/K$^+$ pumps, and also in a very detailed mebrane model that has been 
used intensively for computational studies of spreading depolarizations and seizure--like activity\cite{KAG00,YAO11}.
\section*{Model}

\subsection*{Hodgkin--Huxley (HH) model and reductions}\label{subsec:2.1}
A simple ion--based neuron model can be obtained as a natural extension of the 
Hodgkin--Huxley (HH) model \cite{HOD52}. We list the basic equations of HH that we used for the
sake of completeness, and also comment on two often used model reductions of which one must be 
modified for our study. Furthermore leak currents are specified, which is necessary for the extension towards 
ion--based modeling.

In the HH model, single neuron dynamics is described 
in terms of an electrically active membrane carrying an electric potential $V$, and 
the three gating variables $n$, $m$ and $h$ that render the system excitable. Ion 
species included are sodium, potassium, and an unspecified ion carrying a leak current, 
which can be attributed to chloride in our extended model. The rate equations read \cite{HOD52}:
\begin{eqnarray}
\frac{\mathrm{d}V}{\mathrm{d}t} &=& -\frac{1}{C_m}(I_{Na^+}+I_{K^+} + I_{Cl^-} - I_{app})\ ,	\label{eq:23}\\
\frac{\mathrm{d}n}{\mathrm{d}t} &=& \frac{n_\infty -n}{\tau_n}\ ,								\label{eq:24}\\
\frac{\mathrm{d}h}{\mathrm{d}t} &=& \frac{h_\infty -h}{\tau_h}\ ,								\label{eq:25}\\
\frac{\mathrm{d}m}{\mathrm{d}t} &=& \frac{m_\infty -m}{\tau_m}\ .								\label{eq:26}
\end{eqnarray}
The top equation is simply Kirchhoff's current law for a membrane with capacitance $C_m$ and membrane potential $V$. $I_{app}$ is an externally applied current that may, for example, initiate voltage spikes. The gating variables $n$, $h$, and $m$ are the potassium activator, sodium inactivator, and sodium activator, respectively. Their dynamics is defined by their voltage--dependent asymptotic values $x_\infty$ and relaxation times $\tau_x$ ($x=n$, $m$, $h$). These are given by 
\begin{eqnarray*}
x_\infty 	&=& \frac{\alpha_x}{\alpha_x + \beta_x}\quad\mathrm{and}\\
\tau_x 		&=& \frac{1}{\phi(\alpha_x+\beta_x)}\quad\mathrm{for}\ x=n,\ m\ \mathrm{and}\ h.
\end{eqnarray*}
Here $\phi$ is a common timescale parameter, and the Hodgkin--Huxley exponential functions are
\begin{eqnarray}
\alpha_m	&=& \frac{0{.}1(V+30)}{1 - \exp(-(V+30)/10)}\ , 	\label{eq:27}\\
\beta_m		&=& 4 \exp(-(V+55)/18)\ , 							\label{eq:28}\\
\alpha_n	&=& \frac{0{.}01(V+34)}{1 - \exp(-(V+34)/10)}\ , 	\label{eq:29}\\
\beta_n		&=& 0{.}125 \exp(-(V+44)/80)\ , 					\label{eq:30}\\
\alpha_h	&=& 0{.}07 \exp(-(V + 44)/20))\ ,					\label{eq:31}\\
\beta_h 	&=& \frac{1}{1 + \exp(-0{.}1(V + 14))}\ .			\label{eq:32}
\end{eqnarray}

The individual ion currents read
\begin{eqnarray}
I_{Na^+} &=& (g_{Na}^l + g_{Na}^g m^3h) \cdot (V - E_{Na})\ ,	\label{eq:33}\\
I_{K^+}  &=& (g_K^l + g_K^g n^4) \cdot (V - E_K)\ ,				\label{eq:34}\\
I_{Cl^-} &=&  g_{Cl}^l \cdot (V - E_{Cl})\ ,					\label{eq:35}
\end{eqnarray}
with $g_{ion}^{l,g}$ denoting leak and gated conductances. In fact, Hodgkin and Huxley set up their model with an unspecified leak current and non--leaking sodium and potassium channels. As long as ion dynamics is not considered this is mathematically equivalent to specifying the leak current as being partially sodium, potassium and chloride, but it is physically inconsistent because the reversal potentials for the ions differ. In an ion--based approach, however, the main task of the ion pumps under physiological conditions is to compensate for sodium and potassium leak currents (see next section) while gated currents are extremely small in the equilibrium. So at this point leak currents for all ion species are important. 

The Nernst potentials $E_{ion}$ are given in terms of the ion concentrations $\mathit{[ion]}$ in the intracellular space (ICS) and the extracellular space (ECS) denoted by subscripts $i$ and $e$, respectively:
\begin{eqnarray}
E_{ion}=\frac{26{.}64}{z_{ion}}\ln(\mathit{[ion]}_e/\mathit{[ion]}_i),\label{eq:36}
\end{eqnarray}
for $\mathit{ion}=K$, $\mathit{Na}$, and  $\mathit{Cl}$ and $z_{ion}$ is the ion valence. All model parameters are listed in Tab.~\ref{tab:1}. The units chosen are those typically used and appropriate for the order of magnitude of the respective quantities. Time is measured in $\mathrm{msec}$, potentials in $\mathrm{mV}$, and ion concentrations in $\mathrm{mMol}/l$. The units for conductance densities imply that ionic and pump current densities are in $\mu\mathrm{A}/\mathrm{cm}^2$. For better readability we omit the square brackets on the ion concentrations and simply write $K_{i/e}$, $\mathit{Na}_{i/e}$, and $\mathit{Cl}_{i/e}$.

For $I_{app}=0$ this model is monostable with an equilibrium at $V=-68$ $\mathrm{mV}$. Note that $E_{Na}\neq V$ and $E_K\neq V$ imply that under equilibrium conditions neither $I_{Na^+}$ nor $I_{K^+}$ vanish, but only their sum does. Sufficiently strong current pulses can---depending on their duration---initiate single voltage spikes or spike trains. Constant applied currents can drive the system to a regime of stationary oscillations. The minimal current required for this is usually called rheobase current.

The HH model can be reduced to two dynamical variables in a way that preserves these dynamical features. One common simplification \cite{RIN89} is to eliminate the fastest gating variable $m$ adiabatically and set
\begin{eqnarray}
m=m_\infty(V)\ .\label{eq:37}
\end{eqnarray}
Second, there is an approximate functional relation between $h$ and $n$ that is usually realized as a linear fit \cite{ERM10}. The ion--based model presented in this article, however, contains a stable fixed point with large $n$, and a linear best fit would then lead to a negative $h$. Therefore we will use the following sigmoidal fit to make sure $h$ is non--negative:
\begin{eqnarray}
h = h_{sig}(n) = 1 - \frac{1}{1 + \exp(-6{.}5(n-0{.}35))}\ .\label{eq:38}
\end{eqnarray}
After this reduction the remaining dynamical variables are $V$ and $n$.

\begin{table}[t]
\caption{\label{tab:1} Parameters for Hodgkin--Huxley model}
\begin{tabular}{|l|l|l|}
        \hline
        Name 			& Value \& unit 		& Description \\ \hline
        $C_m$ 			& 1 $\mu$F/cm$^2$ 		& membrane capacitance \\ 
        $\phi$ 			& 3/msec 				& gating timescale parameter\\ 
        $g_{Na}^l$		& 0{.}0175 mS/cm$^2$ 	& sodium leak conductance \\
        $g_{Na}^g$		& 100 mS/cm$^2$ 		& max.\ gated sodium conductance \\
        $g_K^l$			& 0{.}05 mS/cm$^2$ 		& potassium leak conductance \\
        $g_K^g$			& 40 mS/cm$^2$ 			& max.\ gated potassium conductance \\
        $g_{Cl}^l$		& 0{.}05 mS/cm$^2$ 		& chloride leak conductance \\
        $\mathit{Na}_i$	& 27 mMol/$l$ 			& ECS sodium concentration \\
        $\mathit{Na}_e$	& 120 mMol/$l$ 			& ICS sodium concentration \\
        $K_i$			& 130{.}99 mMol/$l$		& ECS potassium concentration \\
        $K_e$			& 4 mMol/$l$			& ICS potassium concentration \\
        $\mathit{Cl}_i$	& 9{.}66 mMol/$l$		& ECS chloride concentration \\
        $\mathit{Cl}_e$	& 124 mMol/$l$			& ICS chloride concentration \\
        $E_{Na}$		& 39{.}74 mV			& sodium Nernst potential \\
        $E_K$			& -92{.}94 mV 			& potassium Nernst potential \\
        $E_{Cl}$		& -68 mV				& chloride Nernst potential \\
        \hline
\end{tabular}
\end{table}

\begin{table}
\caption{\label{tab:2} Model parameters for ion--based model only}
\begin{tabular}{|l|l|l|}
\hline
Name 				& Value \& unit 												& Description \\ \hline
$\omega_i$ 			& 2{.}16 $\mu$m$^3$ 											& volume of ICS \\ 
$\omega_e$			& 0{.}72 $\mu$m$^3$ 											& volume of ECS \\
$F$ 				& 96485 C/Mol	 												& Faraday's constant \\ 
$A_m$ 				& 0{.}922 $\mu$m$^2$ 											& membrane surface \\ 
$\gamma$ 			& 9{.}556e--6 $\frac{\mu\mathrm{m}^2\mathrm{Mol}}{\mathrm{C}}$ 	& conversion factor \\ 
$\rho$ 			& 5{.}25 $\mu$A/cm$^2$ 												& max.\ pump current \\
\hline
\end{tabular}
\end{table}


\subsection*{Minimal ion--based model}\label{sec:2}
While in the original HH model ion concentrations are model parameters, in ion--based modeling 
intra-- and extracellular ion concentrations become dynamical variables, which causes the Nernst 
potentials to be dynamic. The model defined by the rate eqs.~(\ref{eq:23}), (\ref{eq:24}) and contraint eqs.~(\ref{eq:37}), (\ref{eq:38}) can straightforwardly be extended to make ion concentrations 
dynamic since currents induce ion fluxes. However, under those equilibrium conditions found in HH neither 
$I_{K^+}=0$ nor $I_{Na^+}=0$. Hence we need to include ion pumps\cite{END00} to make sure that the rate 
of change in ion concentration inside the cell ($i$) and extracellular ($e$) can vanish in the resting 
state ($\dot{\mathit{Na}}_{i/e}=\dot{K}_{i/e}=\dot{\mathit{Cl}}_{i/e}=0$).

The rate equations for ion concentrations in the intracellular space (ICS) are then
\begin{eqnarray}
\frac{\mathrm{d}\mathit{Na}_i}{\mathrm{d}t}	&=&	-\frac{\gamma}{\omega_i}(I_{Na^+}+3 I_{p})\ ,	\label{eq:8}\\
\frac{\mathrm{d}K_i}{\mathrm{d}t}			&=&	-\frac{\gamma}{\omega_i}(I_{K^+}-2 I_{p})\ ,	\label{eq:9}\\
\frac{\mathrm{d}\mathit{Cl}_i}{\mathrm{d}t}	&=&	+\frac{\gamma}{\omega_i}I_{Cl^-}\ .				\label{eq:10}
\end{eqnarray}
The factor $\gamma$ converts currents to ion fluxes and depends on the membrane surface $A_m$ and Faraday's constant $F$:
\begin{eqnarray}
\gamma = \frac{A_m}{F}\ ,\label{eq:11}
\end{eqnarray}
Dividing the ion fluxes by the ICS volume $\omega_i$ gives the change rates for the ICS ion concentrations. The pump 
current $I_p$ represents the ATP--driven exchange of ICS sodium with potassium from the extracellular space (ECS) at a 
$3/2$--ratio. It increases with the ICS sodium and the ECS potassium concentration. Chloride is not pumped. We are using 
the pump model from \cite{CRE08,BAR11}:
\begin{eqnarray}
I_p(\mathit{Na}_i,K_e)&=&\rho\bigg(1+\exp{\bigg(\frac{25-\mathit{Na}_i}{3}\bigg)}\bigg)^{-1} \nonumber \\ && \bigg(1+\exp{(5.5-K_e)}\bigg)^{-1}\ ,\label{eq:12}
\end{eqnarray}
where $\rho$ is the maximum pump current. As a consequence of mass conservation ion concentrations in the ECS can be computed 
from those in the ICS\cite{CRE09}:
\begin{eqnarray}
\mathit{ion}_e&=&\mathit{ion}_e^{(0)}+\frac{\omega_i}{\omega_e}(\mathit{ion}_i^{(0)}-\mathit{ion}_i)\ ,
\end{eqnarray}
with the ECS volume $\omega_e$. Superscript zero indicates initial values. Since all types of transmembrane currents, i.e., 
also the pumps, must be included in eq.~(\ref{eq:23}) for the membrane potential, we have to add the net pump current $I_p$:
\begin{eqnarray}
\frac{\mathrm{d}V}{\mathrm{d}t} 			&=& -\frac{1}{C_m}(I_{Na^+}+I_{K^+} + I_{Cl^-} + I_p- I_{app})\ .	\label{eq:13}	
\end{eqnarray}
The rate equations for the ion--based model are thus given by eqs.~(\ref{eq:24}), (\ref{eq:8})--(\ref{eq:10}), (\ref{eq:13}). 
These rate equations are complemented by the gating constraints eqs.~(\ref{eq:37}), (\ref{eq:38}) and the mass conservation constraints
\begin{eqnarray}
\mathit{Na}_e	&=&\mathit{Na}_e^{(0)}	+\frac{\omega_i}{\omega_e}(\mathit{Na}_i^{(0)}	-\mathit{Na}_i)	\ ,\label{eq:14}\\
\mathit{K}_e	&=&\mathit{K}_e^{(0)}	+\frac{\omega_i}{\omega_e}(\mathit{K}_i^{(0)}	-\mathit{K}_i)	\ ,\label{eq:15}\\
\mathit{Cl}_e	&=&\mathit{Cl}_e^{(0)}	+\frac{\omega_i}{\omega_e}(\mathit{Cl}_i^{(0)}	-\mathit{Cl}_i)	\ .\label{eq:16}
\end{eqnarray}
Dynamic ion concentration imply that the Nernst potentials in
eqs.~(\ref{eq:33})--(\ref{eq:35}) are now dynamic (see eq.~(\ref{eq:36})). The additional parameters of the ion--based model are listed in
Tab.~\ref{tab:2}. The morphological parameters $A_m$ and $\omega_i$ are taken
from \cite{KAG00}. In cortical ion--based models, the extracellular volume
fraction $f=\omega_e/(\omega_e+\omega_i)$ ranges from $13\%$ in \cite{KAG00} to
$33\%$ in \cite{ZAN11}. In experimental studies, $f$ is about $20\%$, a value
that can increase, for example, in focal cortical dysplasias type II,  a
frequent cause of intractable epilepsy, to $27\%$ \cite{ZAM12} or during sleep
to $32\%$ (the latter only, if we transfer the increase observed in mouse data
to human) \cite{XIE13}. It is important to note that in experimental studies,
the extracellular volume fraction refers to the fraction with respect to the whole tissue,
which includes also the glial syncytium. Assuming  equally sized  neuronal
and glial volume fractions of 40\% each, an experimentally measured value of $20\%$ would in our model, which does
not directly include the volume of the glial syncytium, correspond to $f=0.33$ or 33\%. 
We choose an intermediate value of $25\%$ for $f$,
but address the influence of the volume ratio in Sec.~Results (see Fig.~\ref{fig:5}). We
prefer to give these morphological parameters in the commonly used units which are appropriate to their order of magnitude
rather than unifying all parameters, e.g.\ the cell volume is given in $\mu\mathrm{m}^3$
instead of $l$ which ion concentrations are related to. Consequently $\gamma$ from Tab.~\ref{tab:2} must be multiplied
by a factor of $10$ to correctly convert currents to change rates for ion concentration
in the given units. Because of the extremely small value of $\gamma$
the membrane dynamics, i.e., the dynamics of $V$ and $n$, is five orders of
magnitude faster than the ion dynamics.

A consequence of this large timescale separation is that the system will attain a Donnan equilibrium when the pumps 
break down. The Donnan equilibrium is a thermodynamic equilibrium state (not to be confused with merely a fixed point, 
though it is one) that is reached for ion exchange across a semipermeable membrane. Since we have not 
explicitly included large impermeable anions inside the cell, this is at first surprising. For no applied currents 
and $I_p=0$, the ion rate equations imply that an equilibrium requires all ion currents to vanish. Since conductances 
are strictly positive it follows that all Nernst potentials and the membrane potential must be equal. Ion concentrations 
will then adjust accordingly. However, eqs.~(\ref{eq:8})--(\ref{eq:10}) and (\ref{eq:13}) imply the following constraint 
on the ICS charge concentration $Q_i$: 
\begin{eqnarray} 
\Delta Q_i:= \Delta\mathit{Na}_i+\Delta K_i -\Delta\mathit{Cl}_i=\frac{C_m\gamma}{\omega_i}\Delta V\ , \label{eq:17}
\end{eqnarray} 
where $\Delta$ denotes the difference between the initial and final value of a variable. Since $\gamma$ is very small, changes in ion concentrations must practically satisfy electroneutrality. This condition together with the equality of all Nernst potentials defines the Donnan equilibrium, so we see that it is contained  in our model as the limit case with no pumps and no applied currents. It should be noted that this observation provides a necessary condition for the correctness of biophysical models.

In this extension of the HH model the ion dynamics makes Nernst potentials time--dependent. The simultaneous effect of 
a diffusive and an electrical force acting on a solution of ions is described more accurately by the Goldman--Hodgkin--Katz 
(GHK) equation though. Nevertheless we prefer Nernst currents, because this formulation allows us to use well--established 
conductance parameters so that the model is completely defined by empirically estimated parameters. In Sec.~Results
we will see how GHK currents can be modelled and that the qualitative dynamical behaviour of the system is not affected (see Fig.~\ref{fig:3}).

\section*{Results}

\subsection*{Phase space analysis of ion--based model}\label{sec:3}
\begin{figure}[t!]
\begin{center}
\includegraphics[width=0.975\columnwidth]{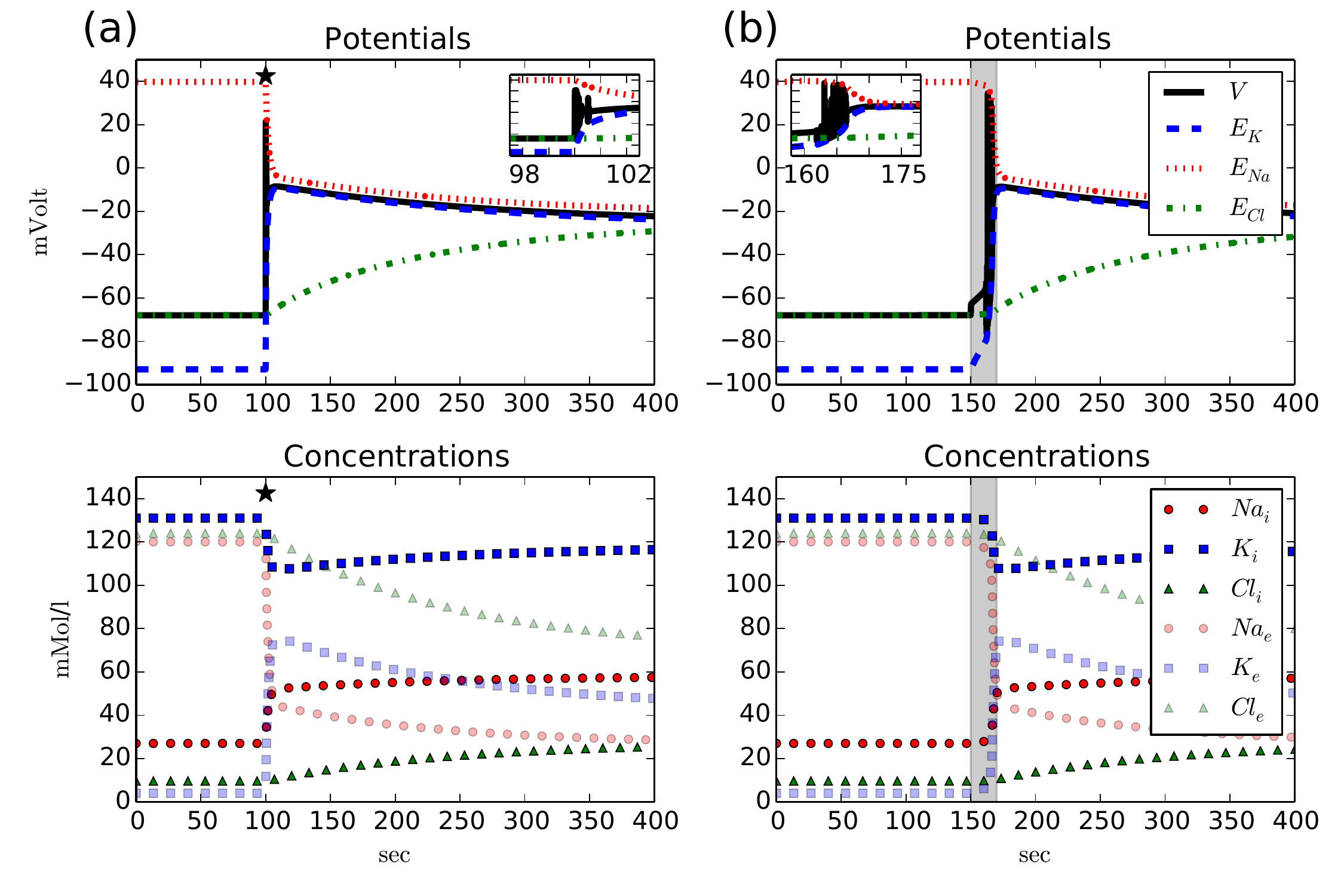}
\end{center} \caption{Upper panels: membrane and Nernst potentials, lower panel: ion concentrations vs time. \textbf{(a)} Response 
of the model to a $0{.}5$ sec long sodium current pulse with amplitude 150 $\mu\mathrm{A}/\mathrm{cm}^2$ (marked by the black star). 
The pulse causes voltage spiking that stops in a strongly depolarized state (see blow--up inset). The membrane potential $V$ takes a 
final value of about $-25\ \mathrm{mV}$ (upper panel). The ion gradients, i.e., the differences between intra-- and extracellular 
ion concentrations, reduce drastically during the stimulation and slowly adjust to a new fixed point after a couple of hundreds of 
seconds (lower panel). \textbf{(b)} Switching off the ion pump for $20\ \mathrm{sec}$ (indicated by the light grey interval) causes 
similar dynamics. The membrane depolarization and dissipation of ion gradients is a bit slower than for (a). After the pump is switched 
on again the system attains the same fixed point as in (a).\label{fig:1}}
\end{figure}

In the ion--based model introduced above current pulses can still initiate voltage spikes (not shown). However, extremely 
strong pulses, in fact comparable to those used in\cite{KAG00} to trigger spreading depolarizations, can drive the system 
away from the physiological equilibrium to a second stable fixed point that is strongly 
depolarized (see Fig.~\ref{fig:1}(a)). This is a new dynamical feature. The depolarized state can also be reached when the 
ion pumps are temporarily switched off (see Fig.~\ref{fig:1}(b)). Apart from the depolarization this state is characterized 
by almost vanishing ion gradients. This free energy--starvation (FES) is reminiscent of the Donnan equilibrium. Extracellular 
potassium is increased from $4$ to more than $40\ \mathrm{mMol/l}$ while the extracellular sodium concentration is reduced 
from $120$ to less than $30\ \mathrm{mMol/l}$. The gated ion channels are mostly open (potassium activation $n$ is $60$\%), 
and it is no longer possible to initiate voltage spikes. In this section we will present a phase space analysis of the model 
and derive conditions for the observed bistability between a physiological equilibrium and a state of FES.

Note that the transition from the physiological state to FES happens via ion accumulation due to spiking, and we will see 
in Sec.~Results that indeed the membrane ability to spike is a necessary condition for the bistability (see Fig.~\ref{fig:3}). Similar 
processes of ion accumulation were regarded as unphysiological in modelling of cardiac cells\cite{DOK93}, but are familiar 
in cortical neurons where ion accumulation is central to seizure--like activity\cite{BAR11,CRE09} and spreading depression\cite{KAG00} 
(SD). In fact, we will briefly demonstrate how the bistability relates to local SD dynamics (see Figs.~\ref{fig:1_new} and \ref{fig:2_new}).

\subsubsection*{Symmetry of the ion--based model}\label{subsec:3.1}
Prior to a bifurcation analysis we need to discuss a conservation law (symmetry) of eqs.~(\ref{eq:8})--(\ref{eq:10}), (\ref{eq:13}). The
direct extension of a membrane model to include ion dynamics as presented above naturally leads to a linear dependence of dynamical variables.
In our case this is reflected by the following relation
\begin{eqnarray}
C_m\frac{\mathrm{d}V}{\mathrm{d}t}=\frac{\omega_i}{\gamma}\left(\frac{\mathrm{d}\mathit{Na}_i}{\mathrm{d}t}+\frac{\mathrm{d}K_i}{\mathrm{d}t}-\frac{\mathrm{d}\mathit{Cl}_i}{\mathrm{d}t}\right)\ ,\label{eq:18}
\end{eqnarray}
for $I_{app}=0$. As a consequence the determinant of the Jacobian is always zero and the system is nowhere hyperbolic. For the 
continuation techniques used by software tools like AUTO\cite{DOE09}, however, the inverse Jacobian plays a central role, so 
they cannot be applied to the system unless this degeneracy is resolved. Furthermore the phase space structure of such nonhyperbolic 
systems can be changed with arbitrarily small perturbations which is why they are called structurally unstable\cite{KUZ95}. Note 
that the linear dependence can be avoided when the rate equation for $V$ contains an additional current with a fixed reversal 
potential breaking the symmetry. Such strictly speaking unphysical currents are indeed often included in neuronal ion--based models\cite{CRE09,BAR11,KAG00,YAO11}, 
but we will rather make use of the symmetry and eliminate one linearly dependent variable.

The physiological view on the instability should be as follows. Assume that the system is in its physiological equilibrium and then 
apply a constant current $I_{app}$ to the voltage rate eq.~(\ref{eq:13}). Then eqs.~(\ref{eq:8})--(\ref{eq:10}) and (\ref{eq:13}) 
imply that the equilibrium conditions $\dot{V}=0$ and $\dot{K}_i=\dot{\mathit{Na}}_i=\dot{\mathit{Cl}}_i=0$ are contradictory, so the 
equilibrium will vanish even for arbitrarily small currents. In fact for any constant and positive $I_{app}$, the system will evolve 
in a highly non--physiological manner with $K_i$, $\mathit{Na}_i$ and $\mathit{Cl}_e$ slowly tending to zero. 

To avoid even the theoretical possibility of such behaviour we will now use eq.~(\ref{eq:18}) to reduce the system and thereby
make it structurally stable. We can, for example, eliminate $V$ and express it in terms of the ICS ion concentrations rather 
than treating $V$ as an independent dynamical variable:
\begin{eqnarray*}
&\frac{\mathrm{d}}{\mathrm{d}t}\left(V-\frac{\omega_i}{C_m\gamma}\left(\mathit{Na}_i+K_i-\mathit{Cl}_i\right)\right)=0\nonumber\\
\Rightarrow \nonumber \\
&V =V^{(0)}+ \nonumber \\  
&\frac{\omega_i}{C_m\gamma}\left(\mathit{Na}_i-\mathit{Na}_i^{(0)}+K_i-K_i^{(0)}-\mathit{Cl}_i+\mathit{Cl}_i^{(0)}\right)
\end{eqnarray*}
This was also done in Ref.\cite{END00}. The physiological meaning of this reduction is simply that the possibility of unspecified applied 
currents is ruled out. For instance, a perturbation on the sodium rate eq.~(\ref{eq:8}) should be interpreted as a sodium current. The above constraint 
describes the simultaneous effect on $V$. It would be equivalent to apply perturbations to eq.~(\ref{eq:8}) and eq.~(\ref{eq:13}) 
consistently to model the full effect of an applied sodium current, so the additional constraint should be seen as a consistency condition. 
(The curves in Fig.~\ref{fig:1}(a) were computed for a sodium current pulse.) This consistency rule does not at all change the dynamics 
unless unspecified currents are applied, and even then it practically does not change the dynamics, because any deviation in ion 
concentrations scales with $\gamma$ and is hence negligible. The structural instability is thus a rather formal feature of the degenerate 
model and we remark that its physiological equilibrium is nevertheless stationary. Instabilities that lead to an unphysiological drift of 
ion concentrations for very long simulation times have been reported and resolved in cardiac cell models\cite{DIF85,DOK93}. Our case is 
different though, because the physiological state is a stationary one and the response to moderate stimulation is physiologically realistic.

For the bifurcation analyses presented in this paper we have eliminated $\mathit{Na}_i$ rather than $V$ for numerical reasons. This is 
completely equivalent, because we only vary the pump rate and morphological parameters. So in our reduction we have replaced rate 
eq.~(\ref{eq:8}) by the following constraint:
\begin{eqnarray}
\mathit{Na}_i &= \mathit{Na}_i^{(0)}-K_i+K_i^{(0)}+\mathit{Cl}_i-\mathit{Cl}_i^{(0)}\nonumber \\ & + \frac{C_m\gamma}{\omega_i}(V-V^{(0)})\label{eq:19}
\end{eqnarray}
The model is then defined by the rate eqs.~(\ref{eq:24}), (\ref{eq:9}), (\ref{eq:10}) and (\ref{eq:13}) and the constraint eqs.~(\ref{eq:37}), 
(\ref{eq:38}), (\ref{eq:14})--(\ref{eq:16}) and (\ref{eq:19}). 

\begin{figure}[t!]
\begin{center}
\includegraphics[width=0.975\columnwidth]{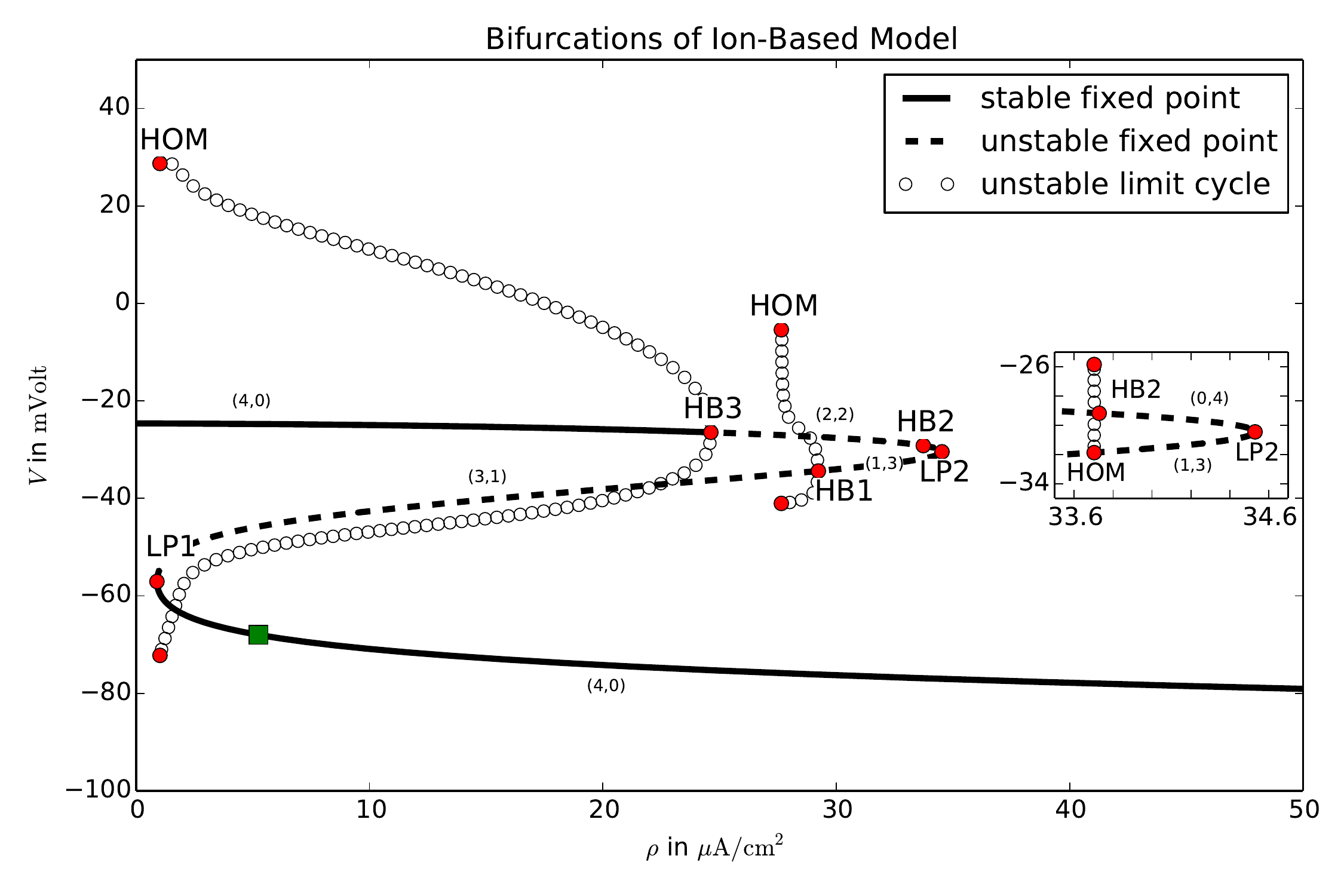}
\end{center} \caption{Bifurcation diagram of the ion--based model. Bifurcations are marked by red circles, the physiological equilibrium by a green square. Following the z--shaped fixed point characteristic from below there are two saddle--node bifurcations (limit point, LP) at $\rho=0{.}894006$ $\mu\mathrm{A/cm}^2$ and $\rho=34{.}5299$ $\mu\mathrm{A/cm}^2$, and three subcritical Hopf bifurcations (HB) at $\rho=29{.}2336$ $\mu\mathrm{A/cm}^2$, $\rho=33{.}7285$ $\mu\mathrm{A/cm}^2$ and $\rho=24{.}6269$ $\mu\mathrm{A/cm}^2$. The limit cycles created in HB1, HB2 and HB3 disappear in homoclinic bifurcations (HOM) at $\rho=27{.}6463$ $\mu\mathrm{A/cm}^2$, $\rho=33{.}7027$ $\mu\mathrm{A/cm}^2$ and $\rho=1{.}024291$ $\mu\mathrm{A/cm}^2$, respectively. The second LP and the second HB together with the HOM of limit cycles occur in a very narrow parameter range (see blow--up inset). The number of stable ($n_-$) and unstable ($n_+$) directions of the fixed point is indicated by the $(n_-,n_+)$--tuples. There is bistability of a physiological state and a depolarized state with largely reduced ion concentration gradients between $\rho=8{.}94006$ $\mu\mathrm{A/cm}^2$ and $\rho=24{.}6269$ $\mu\mathrm{A/cm}^2$.\label{fig:2}}
\end{figure}

\subsubsection*{Bifurcation analysis}\label{subsec:3.2}
We have used the continuation tool AUTO\cite{DOE09} to follow the polarized fixed point of the system under variation of the maximal pump rate $\rho$. Stability changes and the creation of stable or unstable limit cycles are detected by the software which helps us to interpret the dynamical behaviour. For a better overview we will extend our bifurcation analysis even beyond the physiologically relevant range. The full bifurcation diagram is presented in Fig.~\ref{fig:2}.

In the $(\rho,V)$--plane the fixed point continuation yields a smooth z--shaped curve where unstable sections are dashed. The physiological equilibrium is marked by a green square. For higher pump rates the equilibrium remains stable and becomes slightly hyperpolarized. If $\rho$ is decreased the physiological equilibrium collides with a saddle point at $\rho_\mathit{LP1}=0{.}894006$ $\mu\mathrm{A/cm}^2$ in a saddle--node bifurcation (limit point, LP). In a LP the stability of a fixed point changes in one direction (zero--eigenvalue bifurcation). Thus after LP1 the fixed point is a saddle point with one unstable direction. In a Hopf bifurcation (HB) at $\rho_\mathit{HB1}=29{.}2336$ $\mu\mathrm{A/cm}^2$ two more directions become unstable. Via another LP at $\rho_\mathit{LP2}=34{.}5299$ $\mu\mathrm{A/cm}^2$ the last stable direction switches to unstable and the saddle becomes an unstable node. In HBs at $\rho_\mathit{HB2}=33{.}7285$ $\mu\mathrm{A/cm}^2$ and $\rho_\mathit{HB3}=24{.}6269$ $\mu\mathrm{A/cm}^2$ the fixed point becomes a saddle and a stable depolarized focus, respectively. The stability is indicated by the $(n_-,n_+)$--tuples along the fixed point curve with $n_{-,+}$ denoting the number of stable and unstable directions.

In every HB a limit cycle is created. Our model only contains unstable limit cycles that are created in subcritical HBs. In the diagram they are represented by their extremal $V$ values. Such unstable limit cycles are not directly observable, but in the bistable regime they can play a role for the threshold behaviour for the transition from one fixed point to the other. All limit cycles in the model disappear in homoclinic bifurcations (HOM). In a HOM a limit cycle collides with a saddle. When it touches the saddle it becomes a homoclinic cycle of infinite period. After the bifurcation the limit cycle does not exist any more. The limit cycles created in HB1, HB2 and HB3 disappear in HOMs at $\rho=27{.}6463$ $\mu\mathrm{A/cm}^2$, $\rho=33{.}7027$ $\mu\mathrm{A/cm}^2$ and $\rho=1{.}024291$ $\mu\mathrm{A/cm}^2$. The limit cycle emanating from HB1 collides with the upper (i.e., less polarized) saddle, for the other two HOMs the situation is clear, because there is only one saddle available. Since the limit cycles are all unstable these bifurcation details are physiologically irrelevant, but mentioned for completeness.

This bifurcation analysis shows that our model is bistable for a large range of pump rates $\rho_{LP1}<\rho<\rho_{HB3}$.  Strongly depolarized and electrically inactive states of neurons with nearly vanishing ion concentration gradients have been reported in pathological states \cite{DRE11,DRE12}, but in real systems such free energy--starvation (FES) is not stable. In the below section we show how this bistability can be resolved.

\subsubsection*{Ionic excitability}\label{subsec:3.3}
We will now briefly show how the above analyzed model can be modified such that the unphysiological bistability turns into excitability of ion dynamics. For this we follow \cite{BAR11,CRE09} and include an additional regulation term for extracellular potassium. This means that $K_e$ becomes an independent dynamical variable and the constraint eq.~(\ref{eq:15}) must be replaced by its rate equation:
\begin{eqnarray}
\frac{\mathrm{d}K_e}{\mathrm{d}t} = \frac{\gamma}{\omega_e}(I_{K^+}-2 I_{p}) + I_{reg}\ .\label{eq:1_new}
\end{eqnarray}
The regulation term $I_{reg}$ can be interpreted as a diffusive coupling to an extracellular potassium bath or as a phenomenological buffering term. It takes the following form:
\begin{eqnarray}
I_{reg}=\lambda(K_{reg}-K_e)\ , 	\label{eq:2_new}
\end{eqnarray}
where $K_{reg}$ is the potassium concentration of an infinite bath reservoir coupled to the neuron or a characteristic parameter for glial buffering, and $\lambda$ is a rate constant (values given in Tab.~\ref{tab:1_new}). $K_{reg}$ takes values of physiological potassium concentrations and hence stabilizes the physiological equilibrium. This is how $I_{reg}$ regulates ion homeostasis and destabilizes the energy--starved state.
\begin{table}
\caption{\label{tab:1_new} Buffering parameters}
\begin{tabular}{|l|l|l|}
\hline
Name 				& Value \& unit 	& Description \\ \hline
$\lambda$ 			& 2{.}7e--5/msec 	& regulation rate \\ 
$K_{reg}$			& 4 mMol/$l$		& regulation level \\
\hline
\end{tabular}
\end{table}

\begin{figure}[t!]
\begin{center}
\includegraphics[width=0.975\columnwidth]{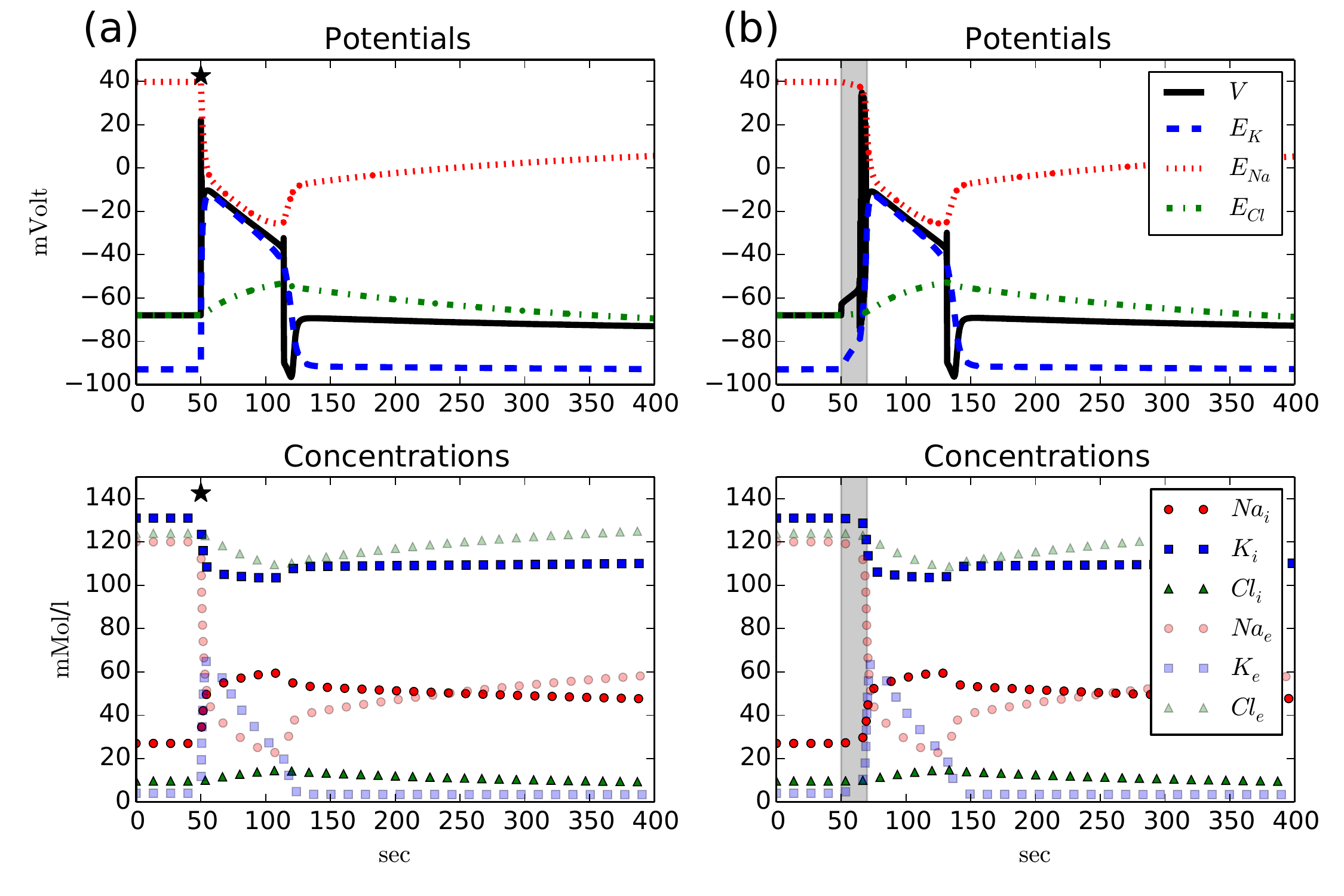}
\end{center} \caption{Upper panels: membrane and Nernst potentials, lower panel: ion concentrations vs time. \textbf{(a)} Stimulation with the same, but earlier applied, current pulse as in Fig.~\ref{fig:1}(a). Due to the additional potassium regulation the system returns to the physiological equilibrium after an approximately 60 sec lasting FES and subsequent hyperpolarization. \textbf{(b)} Similar dynamics as in (a) is observed for a temporary pump switch--off like in Fig.~\ref{fig:1}(b).\label{fig:1_new}}
\end{figure}

If we now stimulate the system with a current pulse or temporarily switch off the pump as we did in Fig.~\ref{fig:1} the system no longer remains in the depolarized state, but repolarizes after a long transient state of FES (see Fig.~\ref{fig:1_new}). After the repolarization ion concentrations start to recover from FES. Full recovery to the initial physiological values is an asymptotic process which takes very long (about two hours), but the neuron is back to normal functioning already after nine to ten minutes. Similar dynamics is described in numerical\cite{KAG00,YAO11} and experimental SD models.

The bistable and excitable dynamics can be nicely compared in a projection of the respective trajectories onto the $(\mathit{Na}_i,K_i)$--plane. For the bistable model the conditions $\dot{K}_i=0$ and $\dot{\mathit{Na}}_i=0$ define three--dimensional hypersurfaces called nullclines. Adding the necessary fixed point conditions on the remaining dynamical variables, namely $V=E_{Cl}$ and $n=n_\infty(V)$, allows us to specify curves that represent these nullclines and only depend on $\mathrm{Na}_i$ and $K_i$. In the buffered model $K_e$ is another dynamical variable and its fixed point condition is $K_e=K_{reg}$. Electroneutrality and mass conservation imply that certain $(\mathit{Na}_i,K_i)$--combinations would lead to a negative $Cl_i$ or $K_e$. In the plot, these unphysiological configurations are shaded.

\begin{figure}[t!]
\begin{center}
\includegraphics[width=0.975\columnwidth]{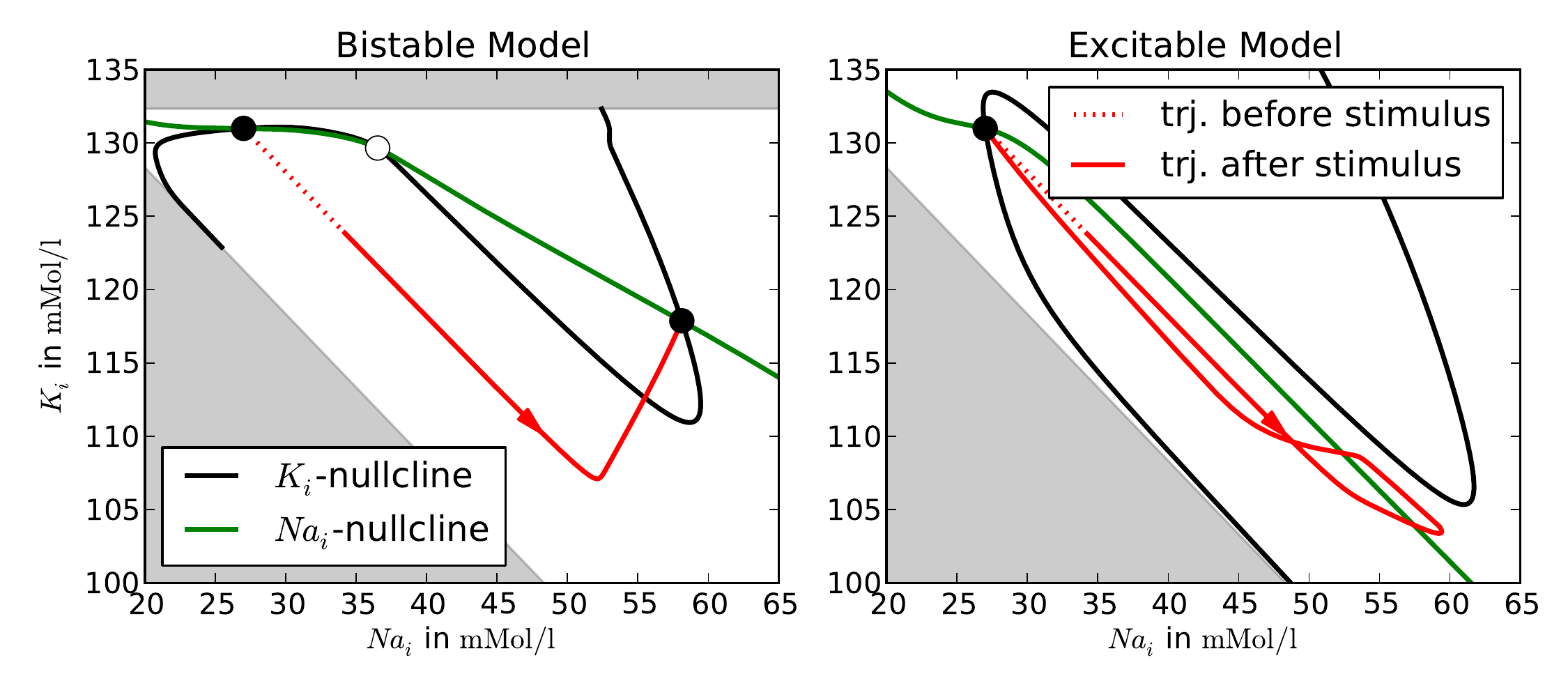}
\end{center} \caption{Projection of the trajectories corresponding to Fig.~\ref{fig:1}a and Fig.~\ref{fig:1_new}a. The shaded regions indicate unphysiological $(\mathit{Na}_i,K_i)$--combinations that imply negative $\mathit{Cl}_i$ (lower left region) or negative $K_e$ (upper region in left plot). Stable and unstable fixed points are marked by solid and open circles. \textbf{(a)} In the bistable case an initial stimulation (dashed line) leads to large subsequent changes in ion concentrations that terminate in the second fixed point of the system. \textbf{(b)} The excitable motion starts very similar to case (a), but after reaching the extremal concentration values the system slowly returns to its initial state. \label{fig:2_new}}
\end{figure}

In Fig.~\ref{fig:2_new}, we see that the bistable model has three nullcline intersections, i.e., fixed points, while the buffering term deforms the nullclines so that only one stable fixed point remains. In the bistable case an initial current pulse stimulation (dashed part of trajectory) drives the system into the basin of attraction of the FES--state, which it then asymptotically approaches (solid line). After the same stimulation the buffered system performs a large excursion in phase space with extremal ion concentrations comparable to FES, but eventually returns to the physiological equilibrium. This large excursion in the ionic variables characterizes what we refer to as ionic excitability or excitability of ion homeostasis. The simulations presented in this section support the hypothesis that it is caused by the bistability of the unbuffered model. Note that intersections of nullcline curves and trajectories do not have to be horizontal or vertical since they may (and do) differ in the non--ionic variables. The main purpose of the nullcline curves is to indicate the existence and location of fixed points. 

\subsection*{Robustness of results}\label{sec:4}
The ion--based model we have analysed so far has been motivated as a natural extension of the Hodgkin--Huxley membrane model. However, there are different variants of ion--based models \cite{KAG00,SHA01,MIU07,SOM08,CRE08,FLO08,BAR11,ZAN11,CHA13} that use different pump and current models, ion content, and ion channels. We will hence address the question how general our results are in this respect. Furthermore we vary the geometry--dependent parameters (membrane surface and extracellular volume fraction) continuously to test their effect on the phase space, too.

\begin{figure}[t!]
\begin{center}
\includegraphics[width=\columnwidth]{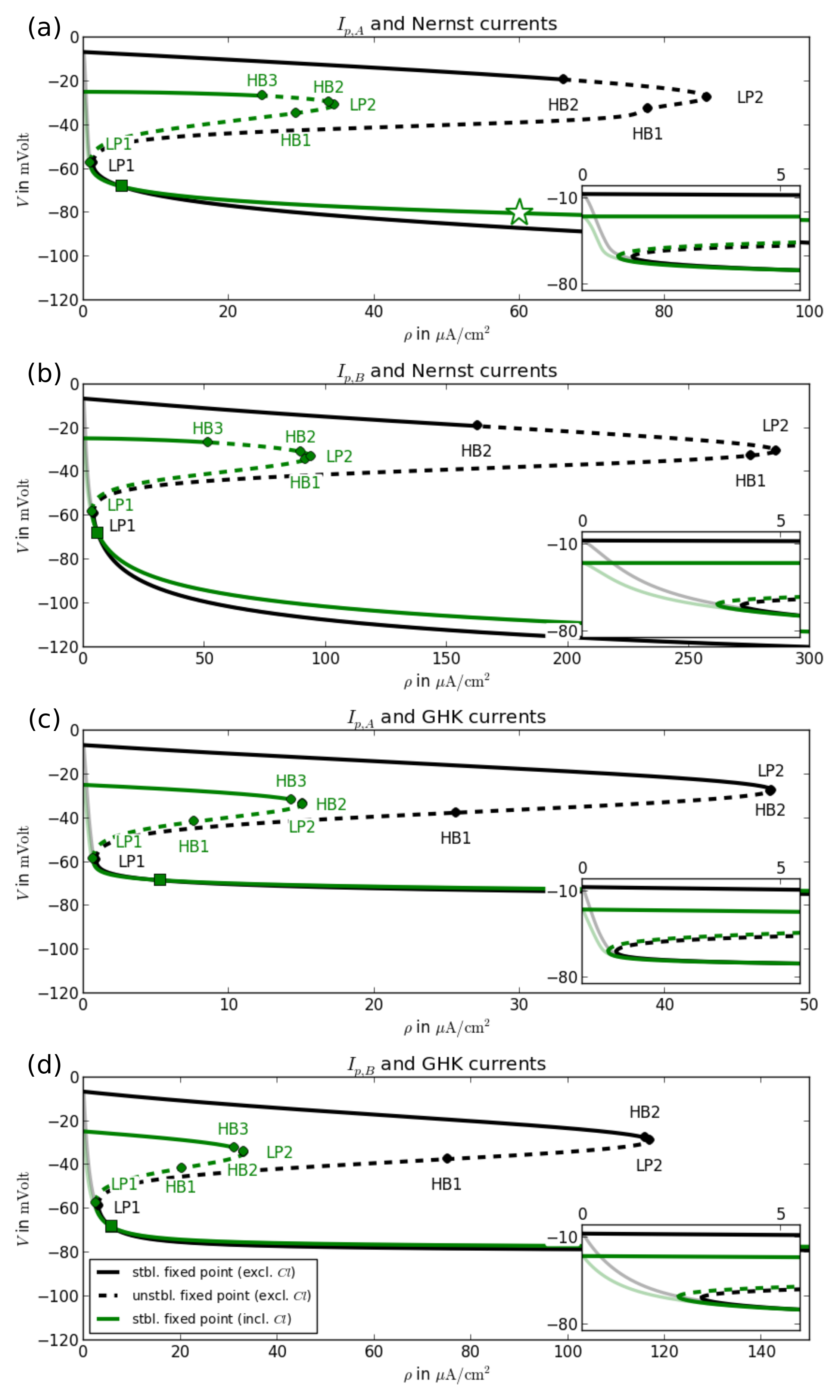}
\end{center} \caption{Bifurcation diagrams of fixed points for different models. The effects of chloride and active ion channels are compared for each of the four possible pump (A vs B) and current model (Nernst vs GHK) combinations. The physiological equilibrium for normal pump rates ($\rho_A=5{.}25$ $\mu\mathrm{A/cm}^2$ and $\rho_B=5{.}72$ $\mu\mathrm{A/cm}^2$) is marked by a green square. The model from fig.~\ref{fig:2} is marked by a star. The value is the same with and without chloride or active channels. Insets show the bifurcation diagrams for low pump rates ($\rho_{A,B}<5$ $\mu\mathrm{A}/\mathrm{cm}^2$). Fixed point lines for models without active ion channels are shaded (see insets). Note the different scales on the main figures, insets are for the same range in each panel.\label{fig:3}}
\end{figure}

\subsubsection*{Model variants}\label{subsec:4.1}
As we noted before, transmembrane currents are more accurately described as Goldman--Hodgkin--Katz (GHK) rather than Nernst currents, even though we prefer the latter. It is hence important to check which difference the choice of current model makes. To generalize the Nernst currents in eqs.~(\ref{eq:33})--(\ref{eq:35}) to GHK currents we assume that both models have the same steady state currents under physiological equilibrium conditions. The GHK version of the sodium current is
\begin{eqnarray}
I_{Na^+}^{GHK} &=& (P_{Na}^l + P_{Na}^gm^3h)\cdot F\cdot\frac{V}{26{.}64}\cdot \nonumber \\ && \frac{\mathit{Na}_i-\mathit{Na}_e\exp(-V/26{.}64)}{1-\exp(-V/26{.}64)}\label{eq:20}
\end{eqnarray} 
with membrane permeabilities $P_{Na}^l$ and $P_{Na}^g$ instead of conductances. To compute these permeabilities we set the GHK current equal to its Nernstian counterpart 
\begin{eqnarray}
I_{Na^+} = (g_{Na}^l + g_{Na}^gm^3h)\cdot(V-E_{Na})\label{eq:21}
\end{eqnarray}
for the equilibrium conditions given in Tab.~\ref{tab:1}. This leads to a common conversion factor from conductances $g_{Na}^{l,g}$ to permeabilities $P_{Na}^{g,l}$. With this ansatz we obtain conversion factors for the three different ion species that lead to the conductances listed in Tab.~\ref{tab:3}.

There is also a certain freedom in the choice of a pump model. It is a general feature of Na$^+$/K$^+$ pumps that their activity is enhanced by the elevation of ECS potassium and ICS sodium. Still different models exist, and to investigate the role of the particular pump model we replace the pump from eq.~(\ref{eq:12}), now referred to as $I_{p,A}$, with the following one from \cite{KAG00}:
\begin{eqnarray}
I_{p,B}(\mathit{Na}_i,K_e)&=& \rho_B \left(1+\frac{3.5}{K_e}\right)^{-2} \left(1+\frac{10}{\mathit{Na}_i}\right)^{-3}\label{eq:22}
\end{eqnarray}
In order to retain the equilibrium at $V=-68$ mV we have to set the maximum pump current to $\rho_B=5{.}72$ $\mu\mathrm{A/cm}^2$. This is slightly higher than the previous pump value ($\rho_A=5{.}25$ $\mu\mathrm{A/cm}^2$), but in the same range.

\begin{table}
\caption{\label{tab:3} Membrane permeabilities for GHK current}
\begin{tabular}{|l|l|l|}
\hline
Name 				& Value \& unit 		& Description					\\ \hline
$P_{Na}^l$ 			& 0{.}0264 $\mu$m/sec 	& leak sodium permeability		\\
$P_{Na}^g$ 			& 150{.}77 $\mu$m/sec 	& gated sodium permeability		\\
$P_K^l$ 			& 0{.}0169 $\mu$m/sec 	& leak potassium permeability	\\
$P_K^g$ 			& 13{.}488 $\mu$m/sec 	& gated potassium permeability	\\
$P_{Cl}^l$ 			& 0{.}0521 $\mu$m/sec 	& leak chloride permeability	\\
\hline
\end{tabular}
\end{table}

\begin{figure}[t!]
\begin{center}
\includegraphics[width=0.975\columnwidth]{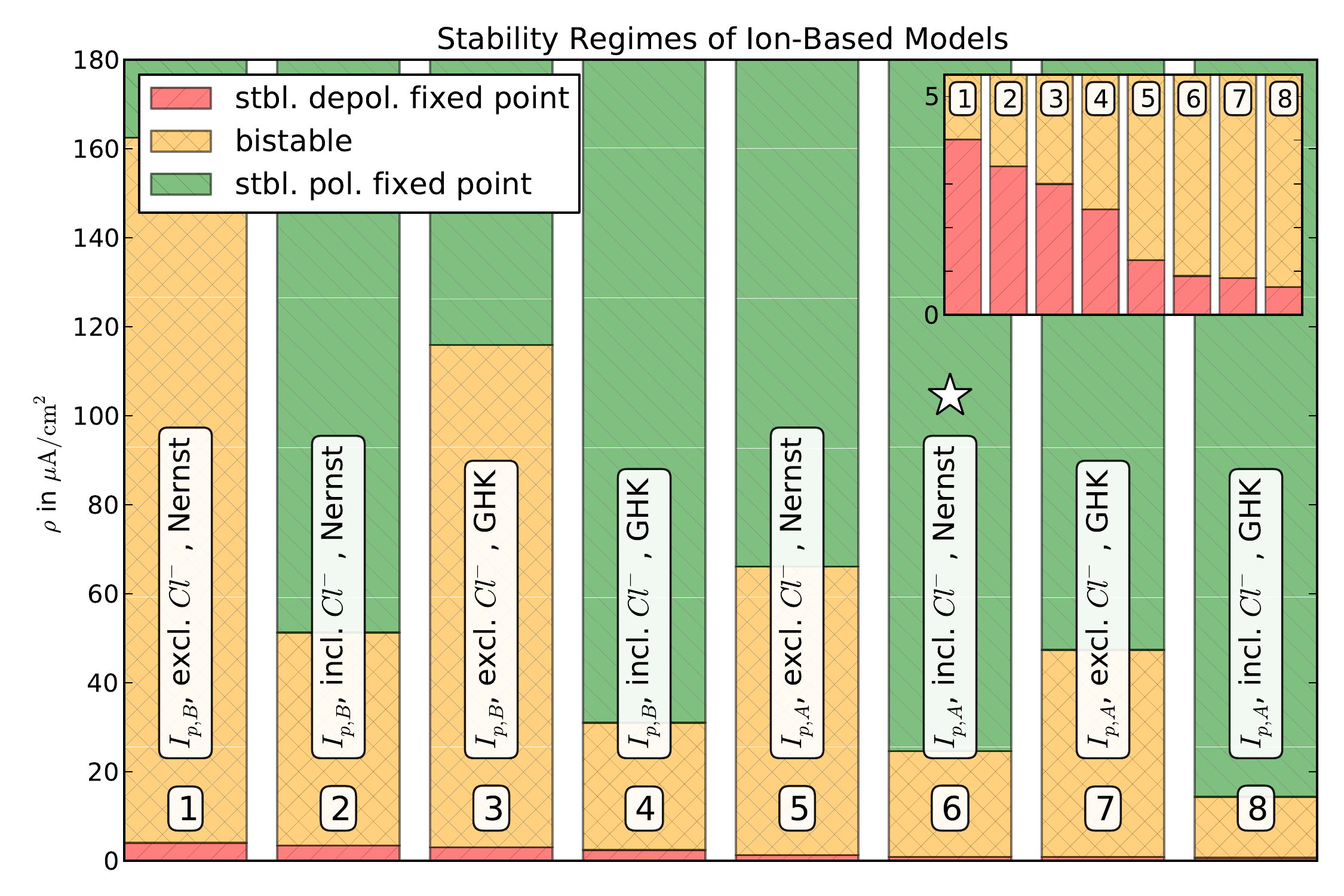}
\end{center} \caption{Overview of the parameter regimes for bistability, polarized and depolarized stability for different models (1--8). The change from the monostable depolarized regime to bistability (red to orange) defines the minimal physiological pump rate, i.e., the pump rate required for the existence of a polarized fixed point. The line separating the bistable from the monostable polarized regime (orange to green) defines the minimal recovery pump rate, i.e., the pump rate required to return from the depolarized fixed point to the polarized equilibrium. The model from Sec.~Model is marked by a star.\label{fig:4}}
\end{figure}

From the rate equations of the HH membrane model (see Sec.~Model) it is obvious that the chloride leak current stabilizes the equilibrium membrane potential. To test its stabilizing effect in the context of ion--based modeling we compare models that either do or do not contain this current. We are further interested in the question whether membrane excitability and ion bistability are related. Therefore also the effect of in-- and excluding active ion channels is tested. 

In this section we will only discuss fixed points and their stability, but not the unstable limit cycles belonging to HBs. In Fig.~\ref{fig:3} the fixed point continuation curves for all combinations of current model (Nernst or GHK), pump choice ($I_{p,A}$ or $I_{p,B}$) and the respective in-- and exclusion of chloride and active ion channels channel are shown. Each panel (a)--(d) contains all continuation curves for a given choice of pump and current model. For those models that are bistable for certain pump rates an overview of the different dynamical regimes is presented in Fig.~\ref{fig:4}. It shows the parameter ranges for bistability and for monostability of a physiological state or FES.

The most striking result of this bifurcation analysis is that this bistability occurs in all models with gated ion channels, but not in any model with only leak channels (grey--shaded graphs in the insets of Fig.~\ref{fig:3}). The comparison of any model with active gates and its leak--only counterpart shows that whenever the physiological equilibrium of the first one exists it is identical to the equilibrium of the latter one. While the physiological state disappears in a LP for all bistable models at small pumping, the fixed points of the leak models remain stable, but depolarize drastically for further decreasing pump rates until the Donnan equilibrium for $\rho_{A,B}=0$ is reached. The absence of the second fixed point in leak--only models is plausible if we consider Fig.~\ref{fig:1} again. The depolarized state is characterized by large ion concentrations $K_e$ and $\mathit{Na}_i$ which implies an increased pump current (see eq.~(\ref{eq:12})). Since the differences between the Nernst potentials and the membrane potential are even smaller in the depolarized state, higher, hence gated, conductances are required to compensate for the pump currents and maintain the depolarized state. Besides the requirement of active ion channels, the bistability is a very robust feature of these simple ion--based models. 

Let us now consider the effect of the different model features on the minimal physiological pump rate, i.e., the pump rate required for a stable physiological fixed point, and the recovery pump rate that destabilizes the depolarized state of FES and allows the neuron to return to physiological conditions. These rates are the lower and upper limit of the bistable regime, and low values are physiologically desirable.

In Fig.~\ref{fig:4} we see that pump model A, GHK currents and chloride each lead to a lower minimal physiological (see the inset) and a lower recovery pump rate than pump model B, Nernst currents and the exclusion of chloride. Quantitative differences should be noted, though. The inset of Fig.~\ref{fig:4} shows that all models with pump A have lower minimal physiological pump rates than models with pump B. So the stability of the physiological equilibrium with respect to pump strength reduction depends mostly on the choice of pump. On the other hand four of the five lowest recovery pump rates are from models that include chloride. In fact, it is only the combination of both, the GHK current model and pump A, that make the recovery threshold of the chloride excluding model 7 slightly lower than that of the chloride including model 2. However, one should note that even the lowest recovery pump rate is as high as $\rho=14{.}3$ $\mu\mathrm{A/cm}^2$ (model 8). This is still an almost threefold increase of the normal rate. So even if we assume pump enhancement due to additional mechanisms, for example increased cerebral blood flow, the threshold for recovery from FES seems to be too high. Thus it is true for a large class of ion--based neuron models that realistic neuronal homeostasis cannot rely on Na$^+$/K$^+$--ATPase alone, but rather on a combination of ion pumps and further regulation mechanisms like glial buffering.

There is another effect of chloride to be pointed out. In Fig.~\ref{fig:3} we see that it raises the Donnan equilibrium potential (see potentials at $\rho=0\ \mu\mathrm{A/cm}^2$) significantly. To understand this effect note that without chloride electroneutrality forces the sum of $\Delta K_e$ and $\Delta\mathit{Na}_e$ to be zero, while the presence of the decreasing anion species $\mathit{Cl}_e$ implies $\Delta(K_e+\mathit{Na}_e)<0$. According to eq.~(\ref{eq:36}) this leads to lower Donnan equilibrium Nernst potentials $E_K$ and $E_{Na}$, and consequently to a lower membrane potential. Since the conditions of FES for physiological pump rate values are very close to the Donnan equilibrium, this depolarized fixed point is shifted in the same way.

The effect of the current model on the two characteristic pump rates is less pronounced than that of chloride or the pump choice. It lowers the minimal physiological rate more than chloride, but not as much as a pump change from model B to A. Its effect on the recovery threshold is the weakest.

\begin{figure}[t!]
\begin{center}
\includegraphics[width=\columnwidth]{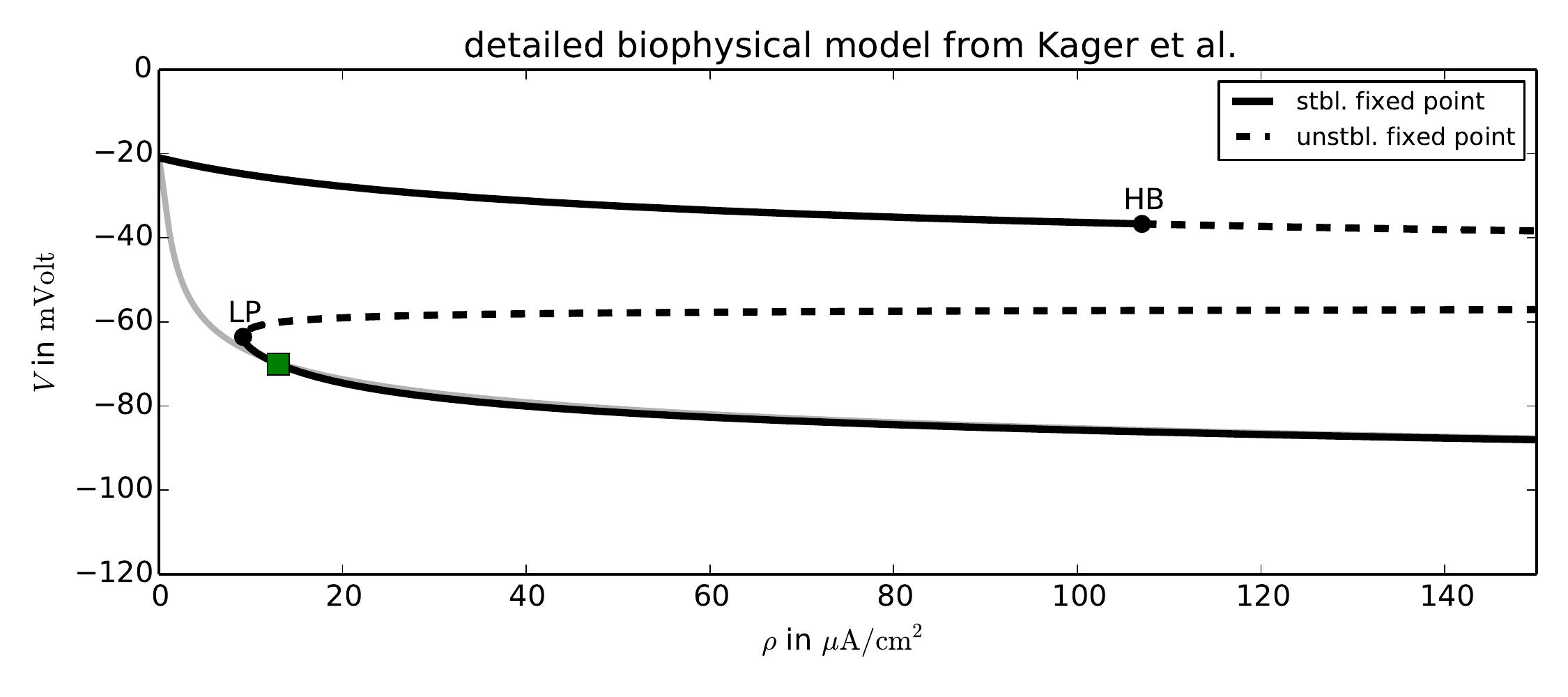}
\end{center} \caption{Bifurcation diagram of the fixed points for the Kager et al.~model\cite{KAG00}. The physiological equilibrium is at $\rho=13$ $\mu\mathrm{A/cm}^2$, the minimal physiological pump rate is $\rho=9{.}8$ $\mu\mathrm{A/cm}^2$, and the recovery rate is $\rho=107$ $\mu\mathrm{A/cm}^2$. Like in Fig.~\ref{fig:3} the fixed point line for the corresponding leak--only model is shaded. Its value at $\rho=0$ $\mu\mathrm{A/cm}^2$ indicates the Donnan equilibrium. \label{fig:3_new}}
\end{figure}

While above we describe and investigate minimal Hodgkin--Huxley model variants to obtain SD behavior in the simplest neuron model types, in the current literature biophysically much more detailed neuron models have been developed for this phenomenon. We do not intend to investigate such detailed models thoroughly, but as an example that further demonstrates the robustness of our results, we also replicate our results with a much more detailed membrane model as first described by Kager et al.\cite{KAG00}. This detailed model contains five different gated ion channels (transient and persistent sodium, delayed rectifier and transient potassium, and NMDA receptor gated currents) and has been used intensively to study spreading depolarizations and seizure--like activity. If we replace the unphysiological fixed leak current of this model, i.e., a leak current with a fixed reversal potential and without associated ion species, with a chloride leak current as in our model and, furthermore, exclude glial buffering, we find the same type of bistability.  The fixed point continuation of this model (for a complete list of rate equations see\cite{KAG00,YAO11}) in Fig.~\ref{fig:3_new} shows that again FES conditions and a physiological state coexist for a large range of pump rates. This model has slightly different leak conductances and equilibrium ion concentrations, and consequently the characteristic pump rates also differ from ours. However, the only important thing to note here is that the recovery pump rate defined by the Hopf bifurcation of the upper fixed point branch is large compared to the physiological value (marked by the green square), so also in this rather different membrane model recovery from FES due to pump enhancement is practically impossible. We remark that the physiologically irrelevant unstable fixed point branches in Fig.~\ref{fig:3_new} do not connect in a saddle--node bifurcation, but saturate for very high pump rates. The occurrence of the same type of bistability in a Hodgkin--Huxley--based ion model and this very detailed one, and also the similarity of the SD trajectories in Fig.~\ref{fig:1_new} to those presented in \cite{KAG00}, support the physiological relevance of the minimal ion--based ansatz that we developed and follow in this paper. Moreover, we assume that bistability is an universal feature also for other more detailed membrane model, which are yet more elaborate variants of the model described by Kager et al.\cite{KAG00}. We will hence use the model from Sec.~Model for further investigations.

\subsubsection*{Variation of membrane surface and extracellular volume fraction}\label{subsec:4.2}
\begin{figure}[t!]
\begin{center}
\includegraphics[width=0.975\columnwidth]{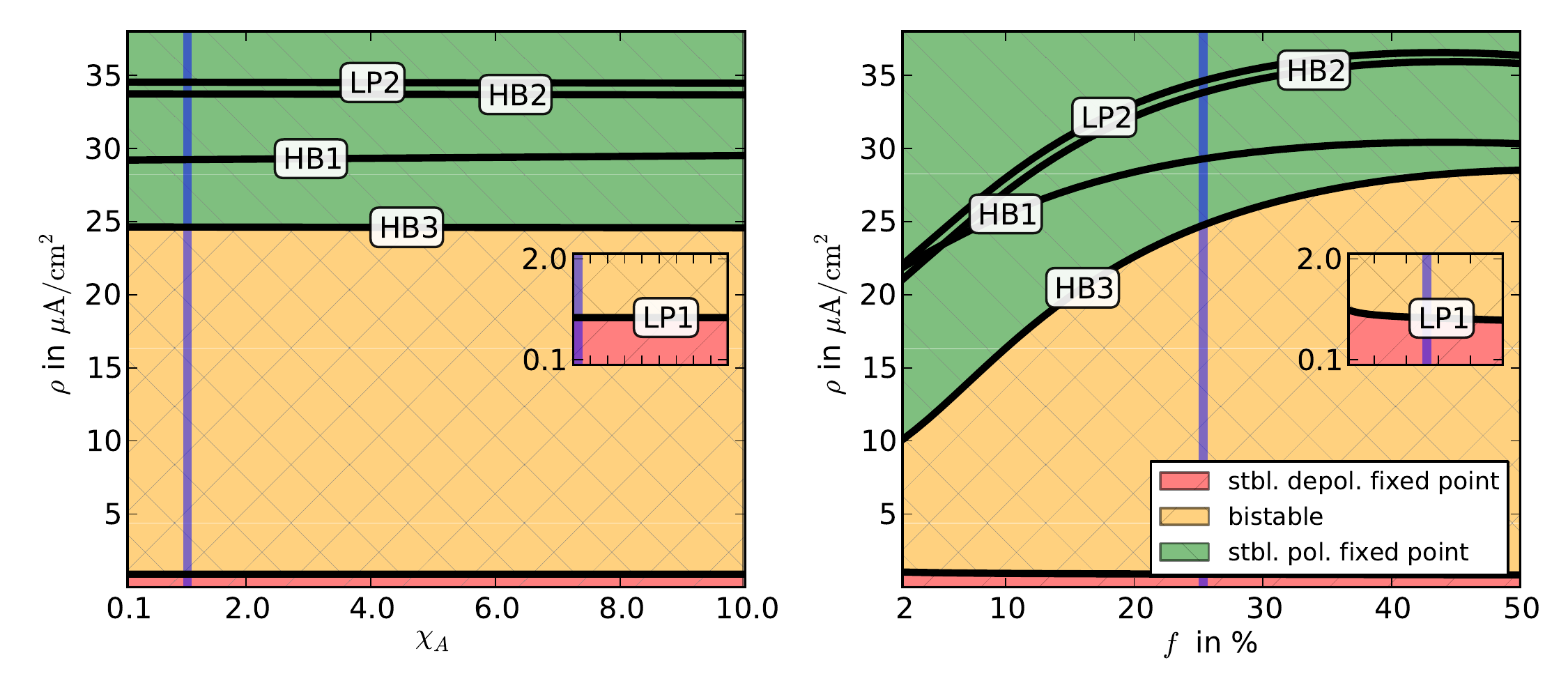}
\end{center} \caption{Two--parameter continuations of the fixed point bifurcations of Fig.~\ref{fig:2}. In the left plot the dimensionless surface size parameter $\chi_A$ is varied, in the right plot the extracellular volume fraction $f$ is changed. The insets show the LP1 curves that mark the minimal physiological pump rate. The pump rates for which the system is bistable range from the LP1 to the HB3. The HB3 pump rate is required to repolarize a neuron that is in the depolarized equilibrium. The parameter $\chi_A$ (left plot plus inset) almost does not change the stability of the system, but $f$ (right plot) reduces the recovery pump rate significantly. The inset shows that the minimal physiological pump rate is much less affected. In each plot and inset the standard parameter value is indicated by the light--blue vertical line.\label{fig:5}}
\end{figure}

After the overview of different variants of ion content, ion channels, pumps and current models we finally address the role of the neuron geometry. Therefore we vary the membrane surface and the extracellular volume fraction in the model from Sec.~Model. For the surface variation we introduce the relative surface size parameter $\chi_A$ and replace $A_m$ with $A_m\chi_A$ which implies the replacement (see eq.~(\ref{eq:11}))
\begin{eqnarray*}
\gamma\rightarrow\gamma\chi_A
\end{eqnarray*}
wherever $\gamma$ occurs, i.e., in the ion rate eqs.~(\ref{eq:9}) and (\ref{eq:10}) and the sodium constraint eq.~(\ref{eq:19}). The extracellular volume fraction, typically denoted as $f$, is defined as
\begin{eqnarray*}
f:=\frac{\omega_e}{\omega_{tot}}\quad\Rightarrow\quad\omega_e=f\omega_{tot}\quad\mathrm{and}\quad\omega_i=(1-f)\omega_{tot}\ ,
\end{eqnarray*}
where $\omega_{tot}=\omega_i+\omega_e$ is the total volume of the system. When $f$ is varied, the above expressions for $\omega_{i/e}$ must be inserted in both ion rate eqs.~(\ref{eq:9}) and (\ref{eq:10}), and in all ion constraint eq.~(\ref{eq:14})--(\ref{eq:16}) and (\ref{eq:19}). The surface parameter $\chi_A$ is varied from $0{.}1$ to 10, $f$  is varied from 2\% to 50\%.  The standard values of these parameters are $\chi_A=1$ and $f=25$\% and parameters are understood to take these values when they are not varied. We start from the bifurcation diagram of Fig.~\ref{fig:2} and perform two--parameter continuations of the detected bifurcations to find out how the membrane surface and the volume ratio change the bistable regime.  The $(\rho,\chi_A)$-- and $(\rho,f)$--continuation curves are shown in the left and right plot of Fig.~\ref{fig:5}.

We see that the $\chi_A$ variation has hardly any effect on the bifurcation values of $\rho$. This can be understood from the structure of the model. The fixed point curve is defined by setting the rate eqs.~(\ref{eq:24}), (\ref{eq:9}), (\ref{eq:10}), (\ref{eq:13}) to zero and the constraint eqs.~(\ref{eq:37}), (\ref{eq:38}), (\ref{eq:14})--(\ref{eq:16}) and (\ref{eq:19}). When $\chi_A$ is varied the only modification to these conditions is in eq.~(\ref{eq:19}) (sodium constraint). But this modification is of order $\mathcal{O}(10^{-5})$ and does practically not affect the shape of the fixed point curve, so the limit point bifurcations LP1 and LP2 are almost not changed. Hopf bifurcation could be shifted, but a rescaling of the (initial and dynamical) ion concentrations by $\chi_A$ transforms the rate and constraint equations such that $\chi_A$ only appears in the pump currents. Their derivatives are then multiplied by $\chi_A$, but for all HBs the pumps are saturated and hence $\chi_A$ does not contribute to the Jacobian. The variation of $f$, however, does change the width (with respect to $\rho$) and the threshold values of the bistable regime. A small value of $f$ (corresponding to a small extracellular space) reduces the recovery pump rate, and also increases the minimal physiological pump rate. This means that both, depolarization and recovery, are enhanced. However, the minimal physiological pump rate is much less affected than the recovery pump rate, so basically a big cell volume supports recovery from the depolarized state. It is known that in spreading depression (SD), where metastable depolarized states that resemble the energy--starved fixed point of our model occur, the osmotic imbalance of ICS and ECS ion concentrations leads to a water influx that makes the cells swell. Our analysis shows that such a process helps the neuron to return to its physiological equilibrium. Extracellular volume fractions of down to 4\% are reported in SD, but even for such extreme volume fractions the required recovery pump rate is too high for pump driven recovery of the neuron (see the lowest value for $\rho_{HB3}$ in the right plot of Fig.~\ref{fig:5}). We remark that the bifurcation curves in Fig.~\ref{fig:5} do not saturate for $f>50$\%, but, except from the LP1--curve that remains very low, bend down, probably due to an approximate symmetry of ICS and ECS ion concentration dynamics. In summary also the analysis of different cell geometries confirms that ion homeostasis cannot be provided by Na$^+$/K$^+$ pumps alone. For example, in computational models of dynamically changing pump rates due to oxygen consumption maximal rates of twice the physiological values are considered\cite{CHA13}.

\section*{Discussion}\label{sec:5}
Computational neuroscience complements experimental and clinical neuroscience.
Simulations help to interpret data and guide a principal understanding of
the nervous systems in both health and disease.  The HH--formulation of
excitability was ``so spectacularly successful that, paradoxically, it created
an unrealistic expectation for its rapid application elsewhere'' as Noble
remarked \cite{NOB01}. While his statement refers to modeling of cardiac cells
it certainly holds true also for neurological diseases and brain
injury\cite{DRE11,DRE12}.  In both fields, the incorporation of the
Na$^+$/K$^+$ pump in the original excitability paradigm  formulated by  Hodgkin
and Huxley is of major importance. The fundamental structure of such models has
to our knowledge not been exploited in neuroscience beyond merely
modulating spiking in epileptiform activity \cite{CRE08,BAR11} or in
models that have energy--starved
states\cite{KAG00,SHA01,MIU07,SOM08,FLO08,ZAN11,CHA13} yet without
investigating the fundamental bifurcation structure.

As we stressed in the introduction, this extension of the original HH model
enforces a physical or rather thermodynamical perspective, which was, of course,
the starting point of Hodgkin and Huxley, too. For instance, we also considered
the Goldman--Hodgkin--Katz (GHK) current equation which is derived from the
constant field assumption applied to the Nernst--Planck equation of
electro\-diffusion. Electroneutrality is important to consider, as can be seen by
the indirect insertion of impermeable counter anions only reflected by
observing a thermodynamic Donnan equilibrium.  Furthermore, a thermodynamic
description of osmotic pressure (which would require a direct insertion of a
concentration $A^{n-}$ of a counter anion with valence $n$) and corresponding
changes in cell volume can be included.

There are further physical mechanisms that may alter the dynamics in
biophysical ion--based models. At the same time, we have to avoid ``an
excruciating abundance of detail in some aspects, whilst other important facets
[...] can only be guessed'' \cite{MAY04}, like using various new currents but
guessing the correct value of the valence $n$ of an impermeable counter anion. For this reason, we decided
to use the original ion currents from the HH model. The comparison of our results to a physilogically more 
realistic and much more detailed membrane model in Fig.\ref{fig:3_new} support the assumption that the basic structure
will not be changed by just adding or modifying gating. This question has
also been addressed experimentally and in simulations by showing that only the
simultaneous blockade of all known major cation inward currents did prevent
hypoxia--induced depolarization with FES\cite{MUL98a,KAG00}.
In the model\cite{KAG00}, five different $\mathit{Na}^+$ currents were
investigated. Of course, to apply our model to a particular
pathological condition, like migraine which is a channelopathy
\cite{SIL13a,SIL13b} (disease caused by modified gating), these details will
become important. This can easily be incorporated in future investigations.
Moreover, note that changes in cell volume, which are very important in brain
injuries, are in this study only treated by varying it as a parameter.

Our bifurcation analysis shows that a whole class of minimal ion--based models
is bistable for a large range of pump rates ($\rho_{LP1}<\rho<\rho_{HB3}$).
Bistable dynamics was suggested by Hodgkin to explain spreading depression\cite{DRE11,CHA13a}, 
and a corresponding model has been investigated mathematically by Huxley
but never been published (cf.\ Ref.\ \cite{DAH04b}). Dahlem and M\"uller suggested to extend this
ad hoc approach, i.e., a single so--called activator variable with a bistable
cubic rate function, by including an inhibitory mechanism in form of an
inhibitor species with a linear rate function coupled to the activator
\cite{DAH04b}.  This, of course, leads to the well known FitzHugh--Nagumo
paradigm of excitability type II\cite{FIT61,NAG62}, that is, excitability caused by a Hopf
bifurcation \cite{ERM98}, but should not be mistaken as a modification of conductance--based
excitability in form of HH--type model in the `first generation' and the
interpretation as an equivalent electrical  circuit. FitzHugh used
his equation in this way, he investigated a long plateau as seen in cardiac
action potentials. Dahlem and M\"uller suggested to use the same mathematical
structure of an activator--inhibitor type model \cite{DAH04b} to describe a
fundamental new physiological mechanism of ionic excitability that originates
from bistable ion dynamics. Our current results provide the missing link
between this ad hoc activator--inhibitor approach, which has been widely used
in migraine and stroke pathophysiology
\cite{DAH04b,DAH07a,DAH08,DAH08d,POS09,DAH09a,SCH09c,DAH12b,POS12,DAH13,DAH13a}, and biophysical
plausible models. The major result from this link is the new interpretation of
the physiological origin of the proposed inhibitory variable
\cite{DAH04b}. We wrongly interpreted it as being related to the pump
rate\cite{DAH07a,DAH08d,DAH09a,DAH13}.

As our ion--based model shows bistable dynamics, we see it as essentially
capturing the activator dynamics of an excitable system and briefly show in
Figs.~\ref{fig:1_new} and \ref{fig:2_new} that it can be transformed into such a  system by the
introduction of a inhibitory process.  Vice versa, excitable systems can be
reduced to bistable dynamics by singular perturbation methods.  Such reductions
are referred to as a threshold reduction. From this perspective our model can
be interpreted as the threshold reduction of an excitable system, and we
conclude that without contact to an ion bath, physically realistic ion--based
models miss an important inhibitory mechanism.  Our analysis shows that
unlike what we thought before\cite{DAH07a,DAH08d,DAH09a,DAH13}, ion pumps alone
are insufficient. If the pump rate is temporarily decreased to less than the
minimal physiological rate, the neuron depolarizes, and normal pump activity
does not suffice to recover the physiological state. Depending on the
particular model the required recovery pump rates range from three times up to
more than 30 times the original value. These high values suggest that also more
detailed pump models that, for example, include the coupling of the maximal
pump rate to oxygen or glucose \cite{CHA13} will not resolve this bistability. 

It can, however, also be seen that a regulation term for the extracellular ion
concentrations that mimics glial buffering and coupling to the vasculature will 
allow only monostability. An additional diffusive coupling to a bath value in the
extracellular rate equations
forces all such buffered extracellular species to assume the respective bath concentrations. 
There are no two points on the solution branch that share the same extracellular potassium
concentrations (see Fig.~\ref{fig:2_new}). Hence one fixed point is selected, the other state 
becomes unstable.  We consequently suspect that coupling to some bath (glia/vasculature) plays a crucial
role in maintaining ion homeostasis and our results from Figs.~\ref{fig:1_new} and \ref{fig:2_new} confirm
that an ion--based model including such coupling will recover from superthreshold
perturbations by a large excursion in phase space that is characterized by long transient free energy--starvation.

\section*{Acknowledgments} The authors are grateful for discussions with Jens
Dreier on spreading depolarizations, scientific guidance by Michael Guevara on
ion--based models in myocytes, and Steven J.\ Schiff and Nancy Kopell, for
various helpful discussions.  


\section*{References}

\end{document}